\documentstyle[preprint,aps,prb,psfig]{revtex}   

\begin{document}

\draft

\title{Charge-transfer metal-insulator transitions in the spin-one-half
Falicov-Kimball model}
\author{Woonki Chung and J. K. Freericks}
\address{
        Department of Physics, Georgetown University,
        Washington, DC 20057-0995
}
\date{\today}
\maketitle

\widetext
\begin{abstract}
The spin-one-half Falicov-Kimball model is solved exactly on an 
infinite-coordination-number Bethe lattice in the thermodynamic limit.
This model is a paradigm for a charge-transfer metal-insulator transition
where the occupancy of localized and delocalized electronic orbitals 
rapidly changes at the metal-insulator transition (rather than the character of
the electronic states changing from insulating to metallic as in a Mott-Hubbard
transition).  The exact solution displays both continuous and discontinuous
(first-order) transitions.
\end{abstract}

\pacs{71.30.+h}

\narrowtext
\section{Introduction}

The Falicov-Kimball Model (FKM)\cite{FKM} is the simplest
model for a charge-transfer metal-insulator transition.  
In this model there are two types of electronic states: (i) localized f-
or d-orbitals
which have a negligible overlap with neighboring electronic orbitals but 
possess a strong on-site Coulomb repulsion (usually taken to be infinite), and 
(ii) delocalized conduction-band orbitals in which the Coulomb repulsion
between two conduction electrons is neglected.  The only ``dynamic''
Coulomb interaction that is included is the Coulomb repulsion between a 
conduction electron and a localized electron that occupy the same lattice site.
The metal-insulator
transition takes place when there is a rapid change in the thermodynamic
occupation of the electron levels as a function of temperature (or some other
thermodynamic variable such as pressure). Hence,
an insulator (or semiconductor) which has most electrons lying in the
localized states rapidly changes it's character to a metal
as the electronic charge transfers from the localized levels to the
conduction band. This transition is similar to the liquid-gas phase 
transition, in which the density changes abruptly at the first-order 
transition temperature or smoothly varies when the temperature is above the 
critical point.  Here it is the density of electrons in the conduction band 
that changes at the transition.

It was originally thought that this model described the most important 
physics behind the metal-insulator transitions in a variety of 
transition-metal and rare-earth compounds.  These materials (such as
SmB$_6$, V$_2$O$_3$, Ti$_2$O$_3$, NiS, etc.) exhibit a variety of
metal-insulator transitions with either continuous or discontinuous 
changes in the conductivity occurring as the temperature is varied.
This opinion, however, was not shared by everyone, since most of these
materials simultaneously exhibit structural and metal-insulator 
phase transitions, and because the FKM ignores 
all effects arising from the hybridization between the localized and
the conduction-band orbitals (such as the Kondo effect, and the screening
of the localized moments).  It remained unclear whether it was the
electronic system, modeled by the FKM, that was driving the transition, or
whether it was driven by other (electron-phonon or hybridization) effects.

Furthermore, there are competing scenarios for metal-insulator transitions.
In the Mott-Hubbard scenario,\cite{mott_hubbard} it is the electron-electron
correlations that change {\it the character of the electrons}
(within a single band) from an insulator to a conductor.
Whereas in the Anderson picture\cite{anderson} it is disorder that 
produces states that are localized and delocalized (within a single band),
and the metal-insulator transition takes place by adjusting the Fermi level
between the localized and delocalized states.  

Recently, however, a series of experiments\cite{nickel_iodide}
have been performed on NiI$_2$,
which appears to be a paradigm for a charge-transfer metal-insulator
transition as described by the FKM.  In this material, the Ni ions donate
one electron to each of the neighboring I ions, filling the Iodine p shell.
The Fermi level lies within the d-band of the Ni ions.  This material is 
an insulator because of the strong electron correlations within the d-band
(which also lead to antiferromagnetic order at low temperatures).  As the
pressure is increased, the N\'eel temperature increases, and then disappears
at the same point where the conductivity has a discontinuous increase by
two orders of
magnitude.  Detailed x-ray diffraction studies showed that there was
no structural phase transition occurring at the metal-insulator transition.
Instead the experimental evidence points toward a transfer of charge from the 
I ions to the Ni ions, which quench the local moments (changing Ni$^{++}$ to
Ni$^+$), and leave behind 
conduction holes in the Iodine p-band.  Such a scenario is described by a 
FKM.\cite{freericks_nii2}

The theoretical side of the FKM has also been controversial.  The original
solution of the model\cite{FKM} involved just a mean-field analysis of
the interband Coulomb interaction.  That approximate
solution found both continuous and discontinuous metal-insulator transitions.
Later, an approximate version of
the Coherent Potential Approximation (CPA) was applied to the model 
by different groups and produced conflicting results:\cite{CPA_yes,CPA_no}
the first-order metal-insulator 
transitions were obtained in one solution;\cite{CPA_yes} while only continuous 
transitions were obtained in the other solution.\cite{CPA_no}  Hence it is
important to be able to solve the FKM in an approximation-free fashion
(and in the thermodynamic limit), to resolve the question of whether or not
it contains discontinuous metal-insulator transitions.

We present an exact solution of the simplest spin-one-half
Falicov-Kimball model on an 
infinite-coordination Bethe lattice (in the thermodynamic limit).  We find 
that the solution possesses both continuous and discontinuous metal-insulator
transitions.  In Section II, we describe the formalism used in solving
the problem, and describe the numerical techniques employed.  In Section III,
we present our results, and end with our conclusions in Section IV.

\section{Formalism}

The FKM consists of two types of electrons: a localized (dispersionless)
valence band and a conduction band
(separated by an energy gap $\Delta$). The Fermi level lies within the
energy gap at zero temperature, so that all of the electrons {\it a priori}
occupy the localized valence band, and the system is an insulator.  Hence, the 
FKM was originally 
described within an electron-hole picture where one considers holes
within the valence band.  The valence holes have a direct Coulomb
repulsion $U_{\tilde{f}\tilde{f}}$, which disfavors two holes occupying the same
localized orbital. In addition, there is an interband on-site
electron-hole Coulomb interaction \mbox{$(-U<0)$} which drives the 
charge-transfer metal-insulator transitions.  The Coulomb interaction is
attractive, because the electron and hole have opposite charge.
The resulting Hamiltonian is 
\begin{equation}
  H = \sum_{{\bf k},\sigma}[\epsilon({\bf k})+\Delta+\frac{W}{2}]
        d^\dagger_{{\bf k},\sigma}d_{{\bf k},\sigma}
      -U\sum_{i,\sigma\sigma^\prime}d^\dagger_{i\sigma}d_{i\sigma}
        \tilde{f}^\dagger_{i\sigma^\prime}\tilde{f}_{i\sigma^\prime}
      +U_{\tilde{f}\tilde{f}}\sum_i
        \tilde{f}^\dagger_{i\uparrow}\tilde{f}_{i\uparrow}
        \tilde{f}^\dagger_{i\downarrow}\tilde{f}_{i\downarrow} \,,
  \label{eq_H_hole}
\end{equation}
where $d^\dagger_{{\bf k},\sigma}$ ($d_{{\bf k},\sigma}$) is the creation 
(annihilation) operator for a conduction electron of wave vector ${\bf k}$ and 
spin $\sigma$, $\epsilon({\bf k})$ is the dispersion of the conduction band  
(with bandwidth $W$), and $\tilde{f}^\dagger_{i\sigma}$ ($\tilde{f}_{i\sigma}$) 
is the creation (annihilation) operator for a localized hole at lattice site 
$i$ (the on-site energy of the localized hole
is chosen as the origin of the energy axis). 
Since $U_{\tilde{f}\tilde{f}}$ is large\cite{FKM} for most localized levels,
we choose the limit where $U_{\tilde{f}\tilde{f}} \rightarrow \infty$ 
and restrict the number of the f-holes per site to 
$n_{\tilde{f}} \leq 1$.  Since the d-electrons in the 
conduction band originate by thermal excitation from the f-band, their number
is constrained to $n_d=n_{\tilde{f}}$ in order to conserve the total number
of electrons. 

We have found it to be more convenient to
study the model in the particle picture rather than 
in the original hole picture given in Eq.~(\ref{eq_H_hole}).  
We employ a particle-hole transformation
($\tilde{f}_{i\sigma} \rightarrow f^\dagger_{i\sigma}$ and 
$\tilde{f}^\dagger_{i\sigma}\tilde{f}_{i\sigma} \rightarrow 
1-f^\dagger_{i\sigma}f_{i\sigma}$) and represent the kinetic energy in the 
localized basis to transform the Hamiltonian into
\begin{eqnarray}
  H &=& -\sum_{ij,\sigma}t_{ij}d^\dagger_{i\sigma}d_{j\sigma}
        +E_f\sum_{i,\sigma}f^\dagger_{i\sigma}f_{i\sigma} 
        -\mu\sum_{i,\sigma}(d^\dagger_{i\sigma}d_{i\sigma} 
          +f^\dagger_{i\sigma}f_{i\sigma}) \nonumber \\
    & & \mbox{}+U\sum_{i,\sigma\sigma^\prime}d^\dagger_{i\sigma}d_{i\sigma}
          f^\dagger_{i\sigma^\prime}f_{i\sigma^\prime}
        +U_{ff}\sum_if^\dagger_{i\uparrow}f_{i\uparrow}
          f^\dagger_{i\downarrow}f_{i\downarrow} \,,
  \label{eq_H_particle}
\end{eqnarray}
where $d^\dagger_{i\sigma}$ ($d_{i\sigma}$) is the creation (annihilation)
operator for a conduction-band electron of spin $\sigma$ at site $i$, 
$t_{ij}$ is the hopping matrix between lattice sites $i$ and $j$ 
[which yields the band structure $\epsilon({\bf k})$],
$E_f=U-\Delta-\frac{W}{2}$ is the localized electron site energy measured 
from the middle of the conduction band, and 
$U_{ff}$ is the on-site Coulomb repulsion between f-electrons.
A chemical potential $\mu$ is introduced to satisfy the constraint 
$n_d+n_f=1$ as $U_{ff} \rightarrow \infty$.  
This Hamiltonian has also been used as a model for intermediate valence 
problems,\cite{valence} 
as a thermodynamic model for an annealed binary alloy,\cite{alloy} 
as a simplified Hubbard model,\cite{simple_hubbard} and as a model for 
metamagnetism and anomalous magnetic response\cite{freericks_zlatic}
(when a magnetic field is added). 

We choose to solve the FKM in the infinite-coordination-number limit 
following the method of exact solution developed by Brandt and 
Mielsch\cite{brandt} and expanded by Freericks and 
Zlati\'c.\cite{freericks_zlatic}
In the infinite-coordination-number limit, the local approximation becomes
exact, implying that one can neglect the momentum dependence of the
irreducible self energy and the irreducible vertex functions, but
one needs to determine the frequency dependence. Hence, the problem on
the lattice can be mapped onto a problem on a single-site, but coupled to
an effective medium, which represents the dynamical information of all of
the other sites of the lattice.  The effective medium needs to be determined
self-consistently in order to solve the lattice problem exactly.

We begin with the local Green's function of the 
conduction electrons for each spin, $G(i\omega_n)\equiv G_n$
evaluated at the Fermionic Matsubara frequencies $\omega_n=(2n+1)\pi k_BT$, 
and express it explicitly in terms of the 
``bare'' Green's function $G_0(i\omega_n)$ (which contains all of 
the dynamical information of the other sites in the lattice):
\begin{equation}
  G_n=\frac{1-n_f}{G_0^{-1}(i\omega_n)}
        +\frac{n_f}{G_0^{-1}(i\omega_n)-U} \ ,
  \label{eq_Gn}
\end{equation}
where 
\begin{equation}
  n_f=\left[1+q\exp{\{(E_f-\mu-U)/k_BT\}}\prod_n
        \{1-UG_0(i\omega_n)\}^{-1}\right]^{-1}.
  \label{eq_nf}
\end{equation}
Here $q=(2J_0+1)/(2J+1)$ is the ratio of the spin degeneracies without 
($J_0$) and with ($J$) an f-electron at a lattice site, and we take 
$q=\frac{1}{2}$. (This is a different value from the original FKM work which 
took $q=2$.  We do not expect the results to depend strongly on the value of 
$q$.) The ``bare'' Green's function also satisfies the usual Dyson equation: 
\begin{equation}
  G_0^{-1}(i\omega_n)=G_n^{-1}-\Sigma_n \,,
  \label{eq_G0}
\end{equation}
which can be viewed as a definition of the self energy 
$\Sigma(i\omega_n)\equiv \Sigma_n$.
The loop for determining the Green's functions
is completed by evaluating the self-consistency equation by summing the 
momentum-dependent Green's function over all momentum [i.e., integrating
over the noninteracting density of states $D(\epsilon)$],
\begin{equation}
  G_n=\int_{-\infty}^{\infty}\!\! d\epsilon \,
        \frac{D(\epsilon)}{i\omega_n+\mu-\epsilon({\bf k})-\Sigma_n}\ ,
  \label{eq_self}
\end{equation}
yielding an exact solution of the model. 

The algorithm for solving the FKM is the same as that used for numerically
solving the Hubbard model:\cite{jarrell_hubbard} (i) Begin with an initial
self energy $\Sigma_n({\rm init.})$ (we chose either 
$\Sigma_n({\rm init.})=0$ or we interpolated from a higher-temperature
run); (ii) use Eq.~(\ref{eq_self}) to find $G_n$; (iii) then use
Eq.~(\ref{eq_G0}) to find $G_0(i\omega_n)$.  (iv) Next
determine $n_f$ from Eq.~(\ref{eq_nf}); and (v) determine the new
$G_n$ from Eq.~(\ref{eq_Gn}).  (vi) Finally use Eq.~(\ref{eq_G0})
with the new Green's function and the old ``bare'' Green's function
to extract the new self energy and (vii) go back to step (ii) to repeat the
iteration until convergence is reached.

Since we are interested in dynamical properties, we also need to solve for 
the retarded Green's functions on the real axis.  We do this by first 
performing an imaginary-axis calculation to find the filling of the 
f-electrons ($n_f$), and then solving the analytically continued equations 
(\ref{eq_Gn}), (\ref{eq_G0}), and (\ref{eq_self}) where the Matsubara 
frequencies are simply replaced by the real frequencies 
($i\omega_n\rightarrow \omega+i\delta$).  
Convergence of these equations is rapid under iteration on both the real and 
imaginary axes (less than one hundred iterations on average for a convergence 
of the self energy to one part in $10^{9}$ on the imaginary axis and to one 
part in $10^{3}$ on the real axis).  

Note that the self-consistent equations for fixed $n_f$, i.e., 
Eqs.~(\ref{eq_Gn}), (\ref{eq_G0}), and (\ref{eq_self}), 
are identical to those employed in the CPA.  
However, $n_f$ is explicitly determined by Eq.~(\ref{eq_nf}) in the 
infinite-coordination-number limit, while, in the CPA, it is  determined  
thermodynamically by minimizing the following trial free energy,
\cite{CPA_yes,CPA_no} $F_{CPA}[n_f]$, as a function of $n_f$:
\begin{eqnarray}
  F_{CPA}[n_f] &=& -2k_{B}T\int_{-\infty}^{\infty}\!\! d\epsilon \,
                     A(\epsilon)\,\ln(1+e^{-\epsilon/k_BT}) \nonumber \\
               & & \mbox{}+(\mu-U+\Delta+\frac{W}{2})(1-n_f) \nonumber \\
               & & \mbox{}+k_{B}T\left[n_f\ln n_f+(1-n_f)\ln \frac{1-n_f}{q}
                   \right] \,,
  \label{eq_F_CPA} 
\end{eqnarray}
which, for the conduction electrons, has the form of a 
noninteracting electron system with the noninteracting density of 
states $D(\epsilon)$ replaced by the interacting one, 
$A(\epsilon)=-\frac{1}{\pi}{\rm Im}G(\epsilon)$. 
Here $\epsilon$ is measured from the chemical potential $\mu$.  
Since the conduction electrons are effectively noninteracting for fixed value 
of $n_f$\cite{van dongen} (i.e., the interacting density of states 
for fixed $n_f$ does
not vary with temperature), the trial free energy in Eq.~(\ref{eq_F_CPA}) 
is also exact.  Naturally, 
the $n_{f}$ obtained in the CPA by minimizing the trial
free energy agrees with the $n_f$ found within the Brandt-Mielsch formalism
using Eq.~(\ref{eq_nf}).  

We solve the model on an infinite-coordination-number Bethe lattice.  
As the coordination number $Z$ increases, the hopping integral $t$ 
between the nearest neighbor sites is scaled as 
$t \rightarrow \frac{t^*}{\sqrt{Z}}$ in order to have a nontrivial kinetic
energy, and the density of states for  the noninteracting system 
becomes Wigner's semicircle with $W=4t^*$:
\begin{equation}
  D(\epsilon)=\frac{1}{2\pi {t^*}^2}\sqrt{4{t^*}^2 - \epsilon^2} \,.
  \label{eq_D}
\end{equation}
We take $t^*$ as our energy unit ($t^*=1$).  
The integral that defines the local Green's function [Eq.~(\ref{eq_self})]
can now be performed analytically, to yield
\begin{equation}
  G(z)=\frac{\tilde z}{2{t^*}^2}-\mbox{sgn}[\mbox{Im}(\tilde z)]
        \frac{\sqrt{\tilde z^2-4{t^*}^2}}{2{t^*}^2}
  \label{eq_gintegral}
\end{equation}
with $\tilde z=z+\mu-\Sigma(z)$ and $z$ an arbitrary complex number.

In addition, one can also calculate susceptibilities for charge-density-wave
order or spin-density-wave order.  When these susceptibilities become infinite
the system has a second-order phase transition to an ordered state.  We
will not describe in detail how to calculate such susceptibilities here
(a discussion has appeared already\cite{brandt,freericks_zlatic}), because
the system never underwent any second-order phase transitions for all of the
parameters we considered in this work.

\section{Results}

We begin our discussion in the low-conduction-band-density limit
$(n_d\rightarrow 0)$ which leads to the ``excitonic phase'' considered by 
Ramirez-Falicov-Kimball.\cite{FKM}
The excitons (bound electron-hole pairs which do not contribute to 
the dc-conductivity) can be shown to form when $U$ is larger than a critical
value $U_c$.  The critical value of $U$ was determined by solving the
single-exciton problem,\cite{FKM} 
where the Hamiltonian in Eq.~(\ref{eq_H_hole}) with one hole and 
one electron produces a bound state just below the 
conduction band.  This occurs when the Green's function, evaluated
on the real axis at the lower band edge, is equal to $-1/U$, or
\begin{equation}
  \frac{1}{U_c}=\int_{-2}^2\!\! d\epsilon \, \frac{D(\epsilon)}{\epsilon+2
    +i\delta} =-G(-2+i\delta)=1 \,,
  \label{eq_U_c}
\end{equation}
for the infinite-coordination-number Bethe lattice after using
Eq.~(\ref{eq_gintegral}).
This is the same critical value of $U$ at which the interacting
density of states splits into two bands as 
$n_d \rightarrow 0$.\cite{van dongen}  
However, it is not clear whether this simple criterion for  
exciton formation (based on the single-exciton problem) is sufficient
to create an excitonic insulator when the conduction-electron density
is small, but finite.
Thus we analytically determine the conduction-band Green's functions at finite 
temperatures and finite, but small electron density.  We find the solutions
separate into two regimes: a weak-coupling regime with $U<1$ and 
a strong-coupling regime with $U>1$. 

The functional form of the self-energy is expressed as a function  
of the local Green's function and the $d$-electron concentration
as:\cite{brandt}
\begin{equation}
  \Sigma(\omega)=\frac{U}{2}-\frac{1}{2G(\omega)}\left\{
        1\pm\sqrt{1+2(1-2n_d)UG(\omega)+U^2G^2(\omega)}\right\} \,,
  \label{eq_Sigma}
\end{equation}
where the sign is chosen for each frequency in such a way that the
self-energy is analytic.  
Since we are interested in the insulating phase ($n_d \rightarrow 0$), we 
expand Eq.~(\ref{eq_Sigma}) about $n_d=0$ by factoring out 
$(1+UG)$ from the square root. If $U<1$ (single-band regime),   the factor
$1+UG(\omega)$ never vanishes [from Eq.~(\ref{eq_U_c})], so we must choose
the negative sign, in order to have a vanishing self energy as $U\rightarrow
0$.  If $U>1$ (two-band regime), we must choose the positive sign for
$\omega < -2$ in order to ensure analyticity (i.e., the right sign for 
Im~$\Sigma$), since $1+UG(\omega)$ is negative there.
Consequently, we choose the minus 
sign for $U<1$ (single band regime) and the plus sign for $U>1$ (two-band 
regime).  Hence, in the limit where 
$\left| \frac{n_dU}{1+UG(\omega)} \right| \ll  1$, we have 
\begin{equation}
  \Sigma(\omega;n_d\rightarrow0)=
  {\everymath{\displaystyle} \left\{ \begin{array}{l}
        U-\frac{n_dU}{1+UG(\omega)}\,,\ \mbox{if $U<1$} \\
        -\frac{1}{G(\omega)}+\frac{n_dU}{1+UG(\omega)}\,,\ \mbox{if $U>1$\,.}
          \rule{0in}{5ex}
  \end{array} \right.}
  \label{eq_Sigma_nd0}
\end{equation}

Combining this with Eq.~(\ref{eq_gintegral}), 
we obtain the functional form of the local Green's function for each case.  
The resulting form is  
\begin{equation}
  G(\omega;n_d\rightarrow0)=
  {\everymath{\displaystyle} \left\{ \begin{array}{l}
        G_{non}(\omega-U)-\frac{n_dUG_{non}(\omega-U)}{
          [1+UG_{non}(\omega-U)]\sqrt{(\omega-U)^2-4}}\,,\ \mbox{if $U<1$} \\
        -\frac{1}{U}+n_dG_{non}(\omega+\frac{1}{U};
          t^*\rightarrow \sqrt{n_d}t^*)\,,\ \mbox{if $U>1$}\,, \rule{0in}{5ex}
  \end{array} \right.}
  \label{eq_G_nd0}
\end{equation}
where we used the noninteracting form of the Green's function, found
in Eq.~(\ref{eq_gintegral}) with $\tilde z\rightarrow \omega +i\delta$.  
Note that
the Green's function for $U>1$ represents only the split-off lower band.

When $U<1$, as expected, the Green's function has just a perturbed form 
from the noninteracting Green's function which is shifted by $U$, 
$G_{non}(\omega-U)$, and the gap to particle-hole excitations at $T=0$ remains 
as $\Delta$.  Thus the number 
of the conduction electrons $n_d$ at finite temperatures is activated  as in
a semiconductor with a fixed energy gap.  
On the other hand, when $U>1$, the Green's function for the lower band becomes 
a band-narrowed ($t^* \rightarrow \sqrt{n_d}t^*$) noninteracting Green's 
function\cite{van dongen} (plus a constant shift by $-\frac{1}{U}$) and has a 
weight $n_d$ (the upper band has weight $1-n_d$).  
The unit-charge condition $(n_f+n_d=1)$ implies that this lower band is always 
half-filled, and the energy of the system (in the limit
$T\rightarrow 0$ and $n_d$ fixed) becomes 
\begin{eqnarray}
  E_{insulator} &\rightarrow& (1-n_d)E_f+
        2\int_{-\frac{1}{U}-2\sqrt{n_d}}^{-\frac{1}{U}}\!\!\!\!\!
        d\epsilon\,\epsilon\,\left[-\frac{1}{\pi}{\rm Im}
        G(\omega;n_d\rightarrow0,U>1)\right] \nonumber \\
  &=&   U-\Delta-2+
        n_d\left(\Delta+2-U-\frac{1}{U}-\frac{8}{3\pi}\sqrt{n_d}\right)\,,
  \label{eq_E_insulator}
\end{eqnarray}
where the factor of 2 in the integral arises from the spin degeneracy of 
the conduction band.
We will see below that this strong-coupling phase is not an excitonic
insulator as believed by Ramirez-Falicov-Kimball\cite{FKM} for any finite
temperature.  
It is interesting to note, though, that this
insulating limit [where all of the electrons lie in the localized
states with an empty conduction band ($n_d=0$)] is noninteracting at $T=0$, 
because there are no conduction electrons, and hence the localized
electrons do not feel any Coulomb repulsion.  The ground-state energy
is $E_{insulator}=U-\Delta-2$.  The gap to particle-hole excitations (at $T=0$
for the insulating phase) is $\Delta$ for $U<1$, but decreases toward zero
as $\Delta+2-U-\frac{1}{U}$ for $U\ge 1$.  At the critical value of $U$ where 
the 
gap closes, one can see from the insulator energy in Eq.~(\ref{eq_E_insulator}) 
that the insulating phase is unstable 
if $\Delta+2-U-\frac{1}{U}<0$ (or equivalently if
$U>1+\frac{\Delta}{2}+\sqrt{\Delta(1+\frac{\Delta}{4})}$) since the ground-state
energy is lowered for small, but nonzero $n_d$.

There is another phase of the FKM that is also noninteracting.
It is the metallic phase (for large $U$) where the
electrons (one per site) in the valence band are all promoted to the
conduction band ($n_d=1$).  In this case, the conduction band feels no Coulomb 
repulsion, because there are no $f$-electrons to scatter them.
Thus the system is characterized by the
half-filled noninteracting conduction band.  The energy of this metallic phase 
is then
\begin{equation}
  E_{metal} = 2\int_{-2}^{0}\!\! d\epsilon \,
                \epsilon \, D(\epsilon) 
            = -\frac{8}{3\pi} \,.
  \label{eq_E_metal}
\end{equation}
Therefore, at $T=0$, there is a transition from an insulating
ground state to a metallic ground state when $E_{metal}<E_{insulator}$ or
\begin{equation}
  U>\Delta+2-\frac{8}{3\pi} \,.
  \label{eq_U_metal}
\end{equation}
Surprisingly, there is a small region of $U$,
$\Delta+2-\frac{8}{3\pi}>U>1+\frac{\Delta}{2}+\sqrt{\Delta(1+\frac{\Delta}{4})}$
and a small region of the gap energy
$0<\Delta<\frac{3\pi}{8}(1-\frac{8}{3\pi})^2\approx 0.026923$, where the 
ground state is {\it neither metallic nor insulating}.  In this nontrivial
region, there must exist either an intermediate-valence state, or 
a charge-density-wave-ordered insulator.  Detailed
studies in this regime will appear in a future publication.

In this contribution, we are interested in examining the discontinuous phase
transitions between states that are connected either to the insulating phase
or to the metallic phase as $T\rightarrow 0$.  So we choose the bare
gap to be large enough ($\Delta=1$), in order to be sufficiently far from
any intermediate valence or charge-density-wave ordered phases.  We vary
$U$ and for each value of $U$ calculate the thermodynamic properties of
the system.  We expect interesting behavior to occur for $U$ close to
the metal-insulator transition point at $T=0$, or $U\approx 2.15117$ 
[from Eq.~(\ref{eq_U_metal})].
We also expect simple semiconducting behavior (with a gap $\Delta$)
to occur for $U<1$.  

In Fig.~\ref{f_rho}, we present our numerical results of $n_d$ (vertical axis)
as a function of $1/T$ (horizontal axis)
for different values of $U$, when $\Delta=1$. The conduction-band filling
$n_d$ is plotted on a
logarithmic scale so that the linear behavior in $1/T$ indicates  
activated carriers as in a semiconductor (see $U=0.5$).  On the other hand, 
the metallic limit obtained in Eq.~(\ref{eq_U_metal}) for $U>2.15117$ also  
agrees with the numerical results, since the system remains metallic for 
all temperatures.  Fig.~\ref{f_rho} shows both discontinuous 
($U=2.150,\,2.140$) and continuous ($U=2.120,\,2.000,\,1.700$) 
metal-insulator transitions.  In the discontinuous transitions, the 
conduction-electron concentration $n_d$ 
follows the metallic solutions at high temperatures but drops to the 
insulating solution at a sharp transition temperature. (For $U=2.150$, 
$n_d$ drops by two orders of magnitude as the temperature changes by
$\sim10^{-3}$.) 
This electron concentration can be viewed as an approximation
to the electrical conductivity, if we assume that in a real material there
is also static disorder (from defects, impurities, etc.) so that at low
temperatures the relaxation time $\tau$ approaches a constant and the 
conductivity is proportional to $n_d\tau$.  We also calculate the {\it 
intrinsic} conductivity of the FKM below, assuming that all of the scattering
of the conduction electrons arises from the localized $f$-electrons.

The conduction-electron density of states, 
$A(\omega)=-\frac{1}{\pi}{\rm Im}G(\omega)$ (where $\omega$ is measured from 
the chemical potential $\mu$) provides additional information about the
metal-insulator transitions.  We calculate it by solving for the Green's
functions on the real-axis.  
Fig.~\ref{f_dos} plots $A(\omega)$ for some of the representative cases from 
Fig.~\ref{f_rho} at various temperatures: (a) the metallic regime ($U=2.160$) 
where the half-filled lower band increases in size as the temperature 
decreases;  (b) the discontinuous metal-insulator-transition regime where
the corresponding density of states in the lower band discontinuously 
collapses to the insulating phase at a critical temperature $T_c$ 
($0.060<T_c<0.065$ for $U=2.150$);  (c) the continuous 
metal-insulator-transition regime where the lower band is continuously 
reduced as the temperature decreases;  (d) and the semiconducting regime where
$n_d$ has an activated behavior, and $A(\omega)$ displays a finite pseudogap 
from the perturbed single conduction band.  Note that all of the temperature
dependence of the interacting density of states arises from the temperature
dependence of $n_d$, since the FKM with fixed $n_d$ and $n_f$ has
a {\it temperature-independent} density of states.\cite{van dongen}

The exact solution of the FKM also allows the optical conductivity to be
calculated from the following formula:\cite{conductivity}
\begin{equation}
  \sigma(\omega)=\sigma_0\int_{-\infty}^{\infty}\!\!\! d\omega'\!
        \int_{-\infty}^{\infty}\!\!\! d\epsilon\, D(\epsilon)\, 
        A(\epsilon,\omega')\, A(\epsilon,\omega'+\omega)\, 
        \frac{f(\omega')-f(\omega'+\omega)}{\omega} \,,
  \label{eq_conductivity}
\end{equation}
where $f(\omega)$ is Fermi distribution function and $\sigma_0$ gives the 
conductivity unit.  Note that this original formula was derived for the
hypercubic lattice.  It relies on two assumptions:  the first is that the
vertex corrections for the conductivity vanish in the infinite-dimensional
limit, and the second is that the derivative of the electronic band-structure
(with respect to the crystal momentum) can be calculated.  The former holds
on a Bethe lattice, but there is no well-defined crystal momentum, and
hence no well-defined bandstructure on a Bethe lattice.  We believe (but have
not been able to show explicitly) that the conductivity for the Bethe lattice
should have the same form as on the hypercubic lattice except for some constants
of order unity (which are absorbed into the definition of $\sigma_0$).
We use this assumption in calculating the dc conductivity $\sigma_{dc}$ 
which is then found from Eq.~(\ref{eq_conductivity}) in the limit 
$\omega\rightarrow0$.

Substituting the spectral function into Eq.~(\ref{eq_conductivity}) and 
assuming the self energy has negligible frequency dependence near
the Fermi level produces the following limiting form for the
dc conductivity with $T=0$ and $n_d\rightarrow0$ (i.e., for the 
insulating phase):
\begin{equation}
  \sigma_{dc}(T=0;n_d\rightarrow0)=-\frac{\sigma_0
        \sqrt{4-[\mu-{\rm Re}\Sigma(0)]^2}}{4\pi^2{\rm Im}\Sigma(0)}\,,
  \label{eq_dc_nd0}
\end{equation}
which is proportional to the product of the {\it intrinsic} relaxation time 
($\sim 1/{\rm Im}\Sigma$) and the density of states at the Fermi surface.  
Calculating the self energy $\Sigma(0)$ from Eqs.~(\ref{eq_Sigma_nd0}) and 
(\ref{eq_G_nd0})  with the proper chemical potential for each case 
($\mu=U-2+\delta\mu\,$ for $U<1$, and $\mu=-\frac{1}{U}$ for $U>1$), we obtain
\begin{equation}
  \sigma_{dc}(T=0;n_d\rightarrow0)=
  {\everymath{\displaystyle} \left\{ \begin{array}{l}
        \frac{\sigma_0(1-U)^2}{2\pi^2U^2n_d}\,,\ \mbox{if $U<1$} \\
        \frac{\sigma_0\sqrt{4-(1+\frac{1}{U})^2}}{
          4\pi^2(U-1)\sqrt{n_d}}\,,\ \mbox{if $U>1$}\,. \rule{0in}{7ex}
  \end{array} \right.}
  \label{eq_dc2_nd0}
\end{equation}
Thus the {\it intrinsic} $\sigma_{dc}$ (which is obtained from the pure 
electronic system of the model) in the insulating phase at $T=0$ 
{\it actually diverges} as $n_d\rightarrow0$!  
Even in the strong-coupling regime ($U>1$), the electron-hole excitations 
do not bind to form an excitonic insulator! (The dc conductivity does
diverge more slowly ($\sim\frac{1}{\sqrt{n_d}}$) than in the weak-coupling 
regime ($\sim\frac{1}{n_d}$) as $n_d\rightarrow0$, though.  )
Therefore, the excitonic phase considered by Ramirez-Falicov-Kimball\cite{FKM} 
does not exist on the infinite-coordination-number Bethe lattice.  
Moreover, the divergence of the {\it intrinsic} $\sigma_{dc}$  
occurs because the relaxation time $\tau$ increases more rapidly 
than the density of states at the Fermi surface decreases when $n_d\rightarrow0$
[see Eq.~(\ref{eq_dc_nd0})].  But in a real material the relaxation time can 
never diverge because there always exists some static disorder 
which forces the relaxation time to approach a constant at low 
temperature. Hence, the dc conductivity will approach zero as 
$n_d\rightarrow0$ in any real material.  
In Fig.~\ref{f_dc}, we present our numerical results of 
the {\it intrinsic} dc conductivity as a function of $1/T$ for the corresponding
values of $U$ in Fig.~\ref{f_rho}.  The conductivity in Fig.~\ref{f_dc}, 
with moderate $n_d$ appears to be proportional to the electron concentration
shown in Fig.~\ref{f_rho}, but 
$\sigma_{dc}$ starts to diverge as temperature decreases and $n_d$ becomes small
enough for the limiting form in Eq.~(\ref{eq_dc2_nd0}) to hold. 
(For example, see the cases $U=2.000$ and $1.700$. For $U=0.500$, 
$\sigma_{dc}$ is always in the low-density limit.  )
Thus, the relevant dc conductivity for a real material  is approximated better
by $n_d\tau$ with a constant relaxation rate $\tau$, rather than using 
the {\it intrinsic} dc conductivity!  

Finally, we examine the discontinuous metal-insulator transitions in 
more detail to show that they are indeed first-order phase transitions.  
We do this by employing the CPA formalism to calculate the trial free energy
as a function of $n_d$ (or equivalently $n_f=1-n_d$) at different 
temperatures.  The minimum of the trial free energy determines the thermodynamic
electron density.  We first solve the real-axis self-consistent equations  for
each value of $0<n_d<1$ to find the interacting density of states
$A(\omega)$. [Note that this $A(\omega)$ at 
fixed $n_d$ is independent of temperature.\cite{van dongen}]  Then we 
evaluate the exact form of the free energy in Eq.~(\ref{eq_F_CPA}) as a 
function of $n_d$ and repeat the process for different temperatures.  
We present our results near the critical temperature in 
Fig.~\ref{f_F_cpa}, where $U=2.150$ and $\Delta=1$.  The free energy 
has a double minimum near the critical temperature ($0.060<T_c<0.065$) and
as the temperature is lowered, the conduction electron density discontinuously
changes as the global minimum switches between the two local minima
(indicating a first-order transition). At $T_c$, where the two minima
are degenerate, the system exhibits phase coexistence between the insulating
and metallic phases.  In the region where the change in $n_d$ is continuous,
the free energy does not have multiple minima, but rather the minimum of
the free energy varies smoothly with $n_d$ as the temperature is changed.
We also verified that the minimum of the CPA form of the free energy agrees
with the form for the free energy determined by 
Brandt and Mielsch.\cite{brandt}

\section{Conclusions}
In conclusion, we have exactly solved the spin-one-half Falicov-Kimball model 
on an infinite-coordination-number Bethe lattice, which is shown to have both 
continuous and discontinuous (first-order) charge-transfer metal-insulator 
transitions.  By being able to solve the model exactly, we have clarified 
the theoretical controversy of the model and have proven that the model 
does display first-order metal-insulator transitions.
The simplicity of the model, based on only the electronic system 
(which has both a localized and a conduction band), 
emphasizes the fact that the electronic system itself 
(the Coulomb interaction between a conduction electron and 
a localized electron) can cause dramatic discontinuous charge-transfer 
metal-insulator transitions without requiring other effects (such as phonons).  
We expect our results to continue to hold in three dimensions and to have
applications to real materials such as NiI$_2$.

We also found that the intrinsic conductivity (determined by the scattering
of the d-electrons off the f-electrons) actually diverges for the ``insulating''
phases with $n_d\rightarrow 0$ because the relaxation time grows faster than
the density of states at the Fermi level decreases.  In a real material,
the conductivity will go to zero as $n_d\rightarrow 0$ though, because the
relaxation time is bounded by the scattering off of impurities.  We also
discovered a small region of parameter space that possesses
intermediate-valence or charge-density-wave order.  Further studies of this
region are currently underway.

\section*{Acknowledgments}
We would like to acknowledge stimulating discussions with 
J. Byers, Ch. Gruber, M. Jarrell, N. Macris, P. van Dongen, and V. Zlati\'c.
This work was supported by the Office of Naval Research Young Investigator
Program under the grant ONR N000149610828.



\newpage
\begin{figure}
  \caption{Number density of the conduction electrons $n_d$, 
(on a logarithmic scale) plotted as a function 
of $t^*/T$ for various values of $U$, where $\Delta=1.0t^*$.  
Four different regimes are shown: (a) the metallic regime for all $T$ 
($U=2.160t^*,\,2.155t^*$); (b) the discontinuous metal-insulator-transition 
regime ($U=2.150t^*,\,2.140t^*$); (c) the continuous metal-insulator-transition 
regime ($U=2.120t^*,\,2.000t^*,\,1.700t^*$); and (d) the semiconducting 
regime ($U=0.500t^*)$.}
  \label{f_rho}
\end{figure}

\begin{figure}[htbp]
  \caption{Density of states $A(\omega)$ at different temperatures for 
some representative cases of Fig.~\ref{f_rho}: 
(a) $U=2.160t^*$; (b) $U=2.150t^*$; (c) $U=2.120t^*$; and (d) $U=0.500t^*$ 
when $\Delta=1.0t^*$.  Here the energy $\omega$ is measured from the chemical 
potential $\mu$ (i.e., the Fermi level lies at $\omega=0$).}
  \label{f_dos}
\end{figure}

\begin{figure}[htbp]
  \caption{The {\it intrinsic} dc-conductivity $\sigma_{dc}$ of the 
Falicov-Kimball model as a function of $t^*/T$ for the corresponding 
values of $U$ shown in Fig.~\ref{f_rho}. Note how the intrinsic
 conductivity diverges for low temperature and low electron concentration, 
as described in the text.}
  \label{f_dc}
\end{figure}

\begin{figure}
  \caption{Free energy as a function of the number density of the conduction 
electrons  near the first-order transition temperature.  The parameters are 
$\Delta=1.0t^*$ and $U=2.150t^*$.  The free energy has two local minima 
with the global minimum switching between these local minima as temperature 
varies through the critical temperature ($0.060<T_c<0.065$), indicating a 
first-order transition.}
  \label{f_F_cpa}
\end{figure}

\newpage
\begin{figure}
  \centerline{\psfig{figure=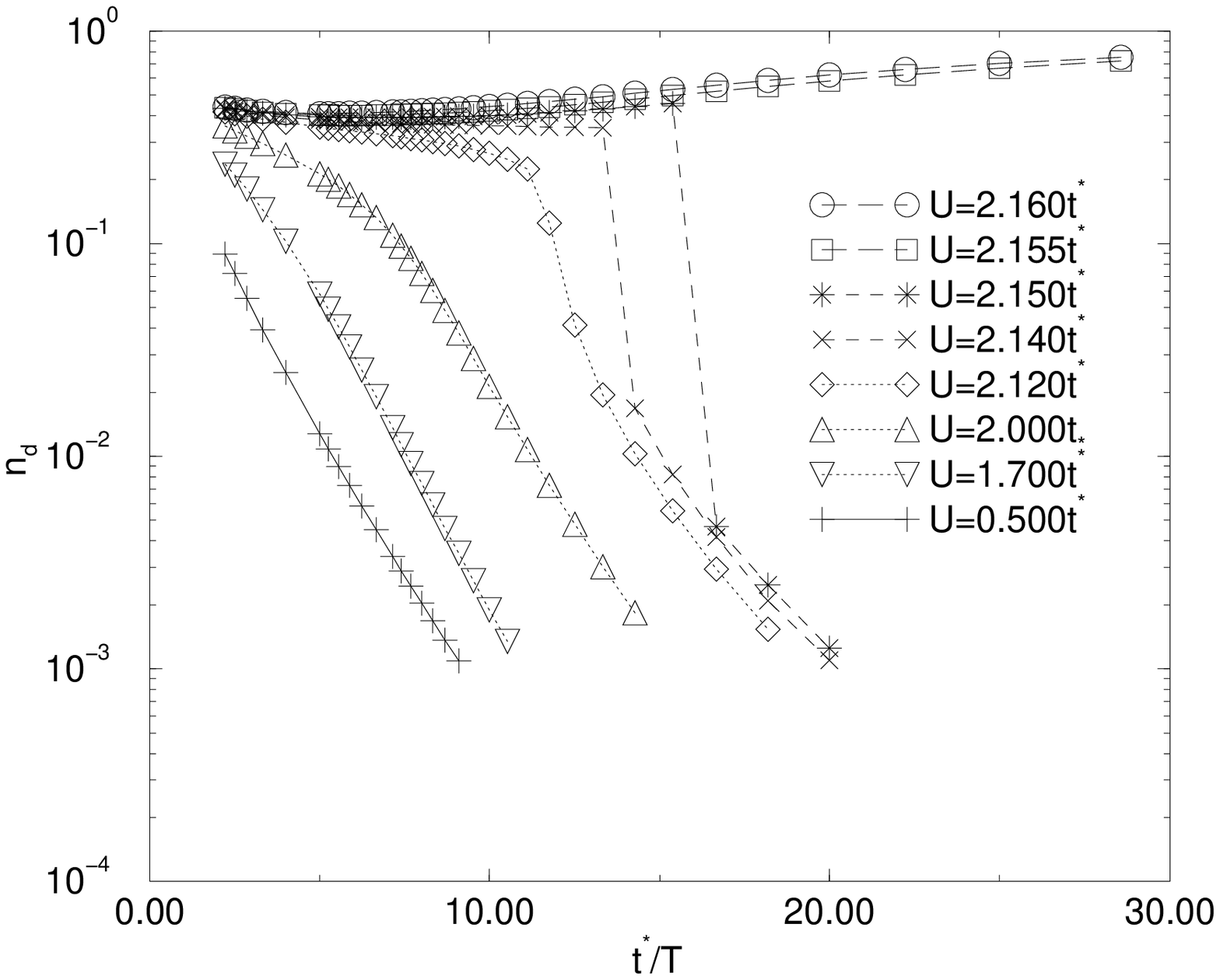,width=6.0in}}
\end{figure}

\newpage
\begin{figure}[htbp]
  \centerline{\psfig{figure=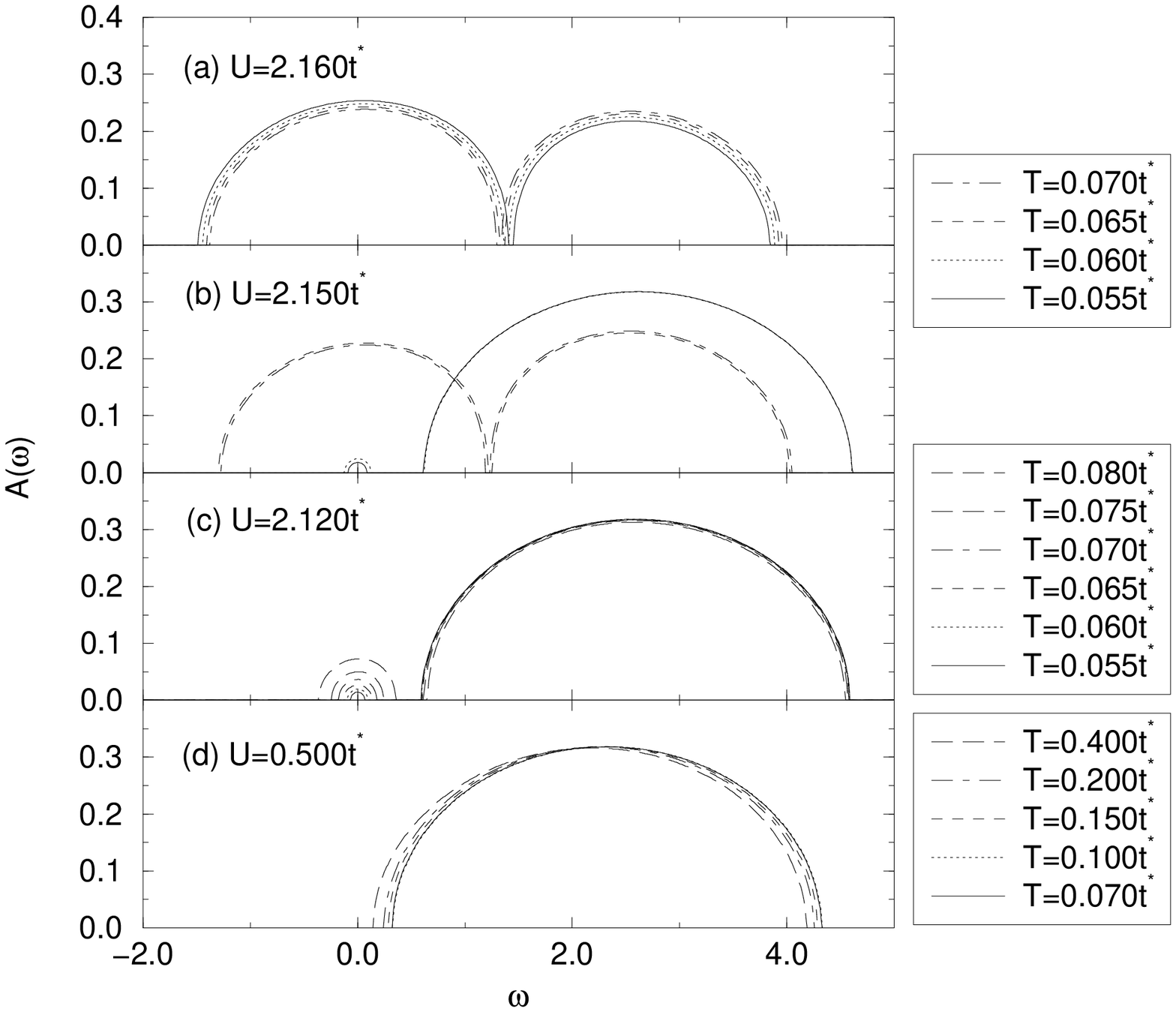,width=6.0in}}
\end{figure}

\newpage
\begin{figure}[htbp]
  \centerline{\psfig{figure=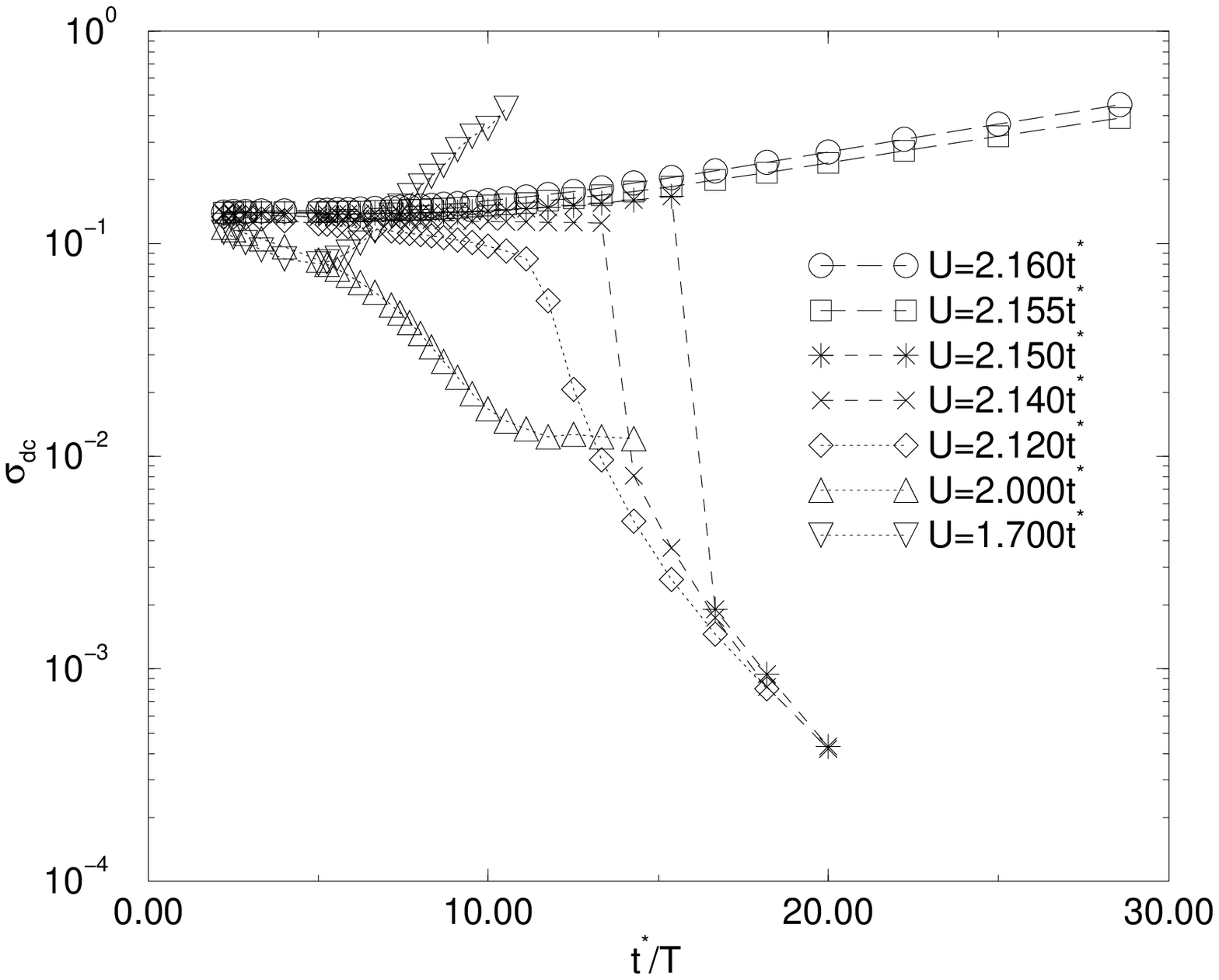,width=6.0in}}
\end{figure}

\newpage
\begin{figure}
  \centerline{\psfig{figure=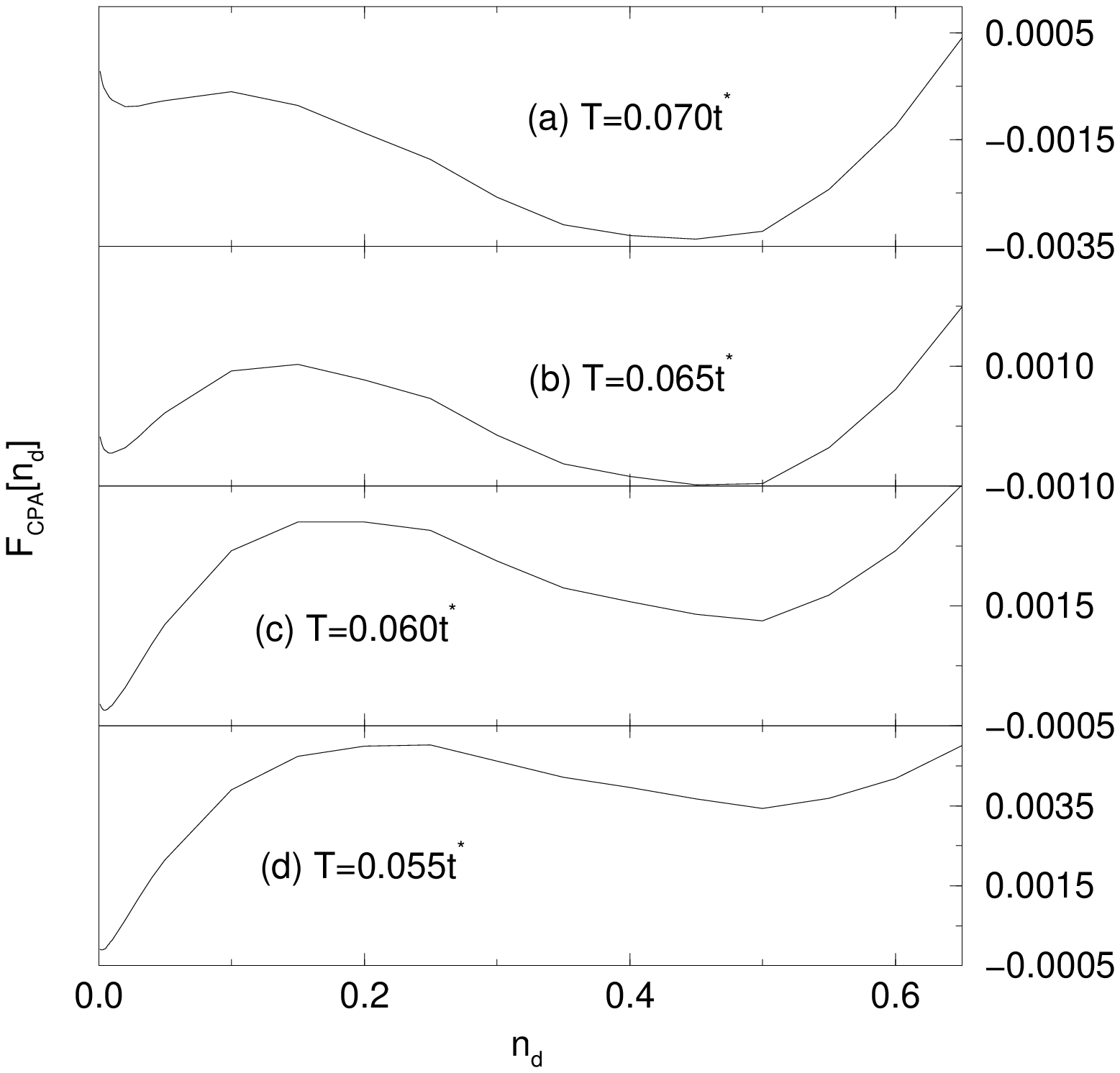,width=6.0in}}
\end{figure}

\end{document}